%Paper: hep-th/9403031
%From: khuri@nxth04.cern.ch
%Date: Fri, 4 Mar 94 15:48:49 +0100
%Date (revised): Fri, 4 Mar 94 16:25:43 +0100
%Date (revised): Fri, 4 Mar 94 17:43:46 +0100

%Paper: hep-th/9403031
%From: khuri@nxth04.cern.ch
%Date: Fri, 4 Mar 94 15:48:49 +0100
%Date (revised): Fri, 4 Mar 94 16:25:43 +0100

\magnification=1200
\hoffset=-.1in
\voffset=-.2in

\vsize=7.5in
\hsize=5.6in
\tolerance 10000

\baselineskip 12pt plus 1pt minus 1pt
\def\footnoterule{\kern-3pt \hrule width \hsize \kern6.2pt}
\def\pmb#1{\setbox0=\hbox{$#1$}%
\kern-.025em\copy0\kern-\wd0
\kern.05em\copy0\kern-\wd0
\kern-.025em\raise.0433em\box0 }

\def\IR{{I\kern-0.25em R}}

\pageno=0
\footline={\ifnum\pageno>0 \hss --\folio-- \hss \else\fi}
%\centerline{\bf }%title
%\smallskip
\hfil\hbox{CERN-TH.7187/94}
\centerline{}
\vskip 24pt
\centerline{{\bf The Wu-Yang Ambiguity Revisited}\footnote{*}{This
work is
supported in part by funds provided by the National Science
Foundation (NSF) Grant PHY/9206867  and the U. S. Department of
Energy (DOE) under contract
\#DE-AC02-76ER03069.}}
\vskip 24pt
\centerline{Daniel Z.~Freedman\footnote{$^1$}{Permanent address:
Department of Mathematics and Center for Theoretical Physics,
Massachusetts
Institute of Technology, Cambridge, MA\ \ 02139-4307, USA.} and
Ramzi R.~Khuri\footnote{$^2$}{Supported by a World Laboratory
Fellowship.}}
\medskip
\vskip 12pt
\centerline{\it Theory Division}
\centerline{\it CERN}
\centerline{\it CH-1211 Geneva 23}
\centerline{\it Switzerland}
\vskip 1.0in
\baselineskip 24pt plus 2pt minus 2pt
\centerline{\bf ABSTRACT}
\medskip
Several examples are given of continuous families of SU(2) vector
potentials
$A_i^a(x)$ in 3 space dimensions which generate the same magnetic
field
$B^{ai}(x)$ (with det $B\neq 0$).  These Wu-Yang families are
obtained from
the Einstein equation $R_{ij}=-2G_{ij}$ derived recently via a local
map of the gauge field system into a spatial geometry with $2$-tensor
$G_{ij}=B^a{}_i B^a{}_j\det B$ and connection $\Gamma_{jk}^i$ with
torsion
defined from gauge covariant derivatives of $B$.

\vfill
\bigskip
\centerline{Submitted to: {\it Physics Letters B}}
\vfill
\vbox{\baselineskip12pt\hbox{CERN-TH.7187/94}
\hbox{CTP\#2282}
\hbox{hep-th 9403031}
\hbox{March 1994}}
\eject
\baselineskip 13pt plus 1pt minus 1pt

The Wu-Yang ambiguity [1] is the phenomenon that two or more gauge
inequivalent non-abelian potentials $A_i^a(x)$ generate the same
field
strength $F_{ij}^a(x)$.  This was widely discussed in the literature
[2-14] of
the 1970's.  Although the original example is 3-dimensional, it was
mainly the
4-dimensional case which was of past interest.  Many examples of a
discrete
ambiguity have been exhibited, specifically two potentials $A$ and
$\bar A$
giving the same $F$.  The few examples of a continuous ambiguity
[7-14] were
degenerate in some way: for example, they were effectively
2-dimensional.

In this communication we study the Wu-Yang ambiguity of SU(2) gauge
field
configurations in 3-spatial dimensions, and we obtain several
examples of
continuous families of potentials which generate the same magnetic
field
$$
B^{ai}=\epsilon^{ijk}\left[\partial_j A_k^a+{1\over 2}\epsilon^{abc}
A_j^b
A_k^c\right]\ .
\eqno(1)$$
In 3 dimensions there is no ``algebraic obstruction'' to an
ambiguity.  Given
two potentials, $A$ and $\bar A$, we find by subtraction of their
Bianchi
identities that their difference $\Delta A=A-{\bar A}$ satisfies
$$
\epsilon^{ijk}\epsilon^{abc} B^{ki}\Delta A_i^c=0
\eqno(2)$$
which is an underdetermined linear system --- 3 equations for 9
unknowns ---
for any $B$-field configuration.  In four dimensions one gets a
``square''
homogeneous linear system --- 12 equations for 12 unknowns --- which
has a
non-trivial solution only if the field strength satisfies an
algebraic
condition corresponding to vanishing determinant [5,6,8].  In any
case, (2) is not
sufficient to demonstrate that (1), viewed as a partial differential
equation
for $A_j^b$ given $B^{ai}$, has multiple solutions, and it is this
which we
wish to explore here, largely by means of examples.

The 3-dimensional case is relevant for the Hamiltonian form of gauge
field
dynamics in $3+1$ dimensions and especially for attempts [15], [16]
to transform
from $A_i^a$ to $B^{ai}$ as the basic field variables. In fact the
work here
emerged from further study of the recent proposal [16] to do this as
an
intermediate step to replace the $A$, $B$ system by a set of
gauge-invariant
spatial geometric variables, namely a metric $G_{ij}$ and connection
$\Gamma_{ij}^k$ with torsion.  It turns out that the information we
find on
the Wu-Yang ambiguity invalidates the form of Hamiltonian dynamics
proposed
in [16].  But the geometry described there is valid, and it is
through the
geometrical equations that our Wu-Yang information is obtained. See
[17] for an
earlier and similar geometrical treatment of the equations of motion
of 2+1 dimensional $SU(2)$ Yang-Mills theory.

1.\quad We begin with our first example.  Consider the smooth,
algebraically
non-singular ({\it i.e.\/} det $B\neq0$) magnetic field
$B^{ai}=\delta^{ai}$,
in Euclidean space with Cartesian coordinates $x,y,z$. It is
easy to show explicitly that, for any real parameter $\beta$ with
$|\beta|>1$,
the 1-parameter family of potentials
$$
A_i^a=\pmatrix{\beta\pm\sqrt{\beta^2-1}
                 \cos (z/\beta) & \pm\sqrt{\beta^2-1}\sin
                 (z/\beta) & 0 \cr
               \pm\sqrt{\beta^2-1}
                 \sin (z/\beta) & \beta\mp\sqrt{\beta^2-1}\cos
		 (z/\beta) & 0 \cr
	       0 & 0 & 1/\beta \cr}
\eqno(3)$$
all reproduce the same $B^{ai}$. Gauge inequivalence is demonstrated
by the
fact that the invariants $B^{aj}D_i B^{ak}$ depend on $\beta$ and
$z$.

\item{a.} Note the ``nonlinear wave'' character of these potentials
--- the
amplitude and wavelength of the trigonometric terms are correlated.

\item{b.} The particular magnetic field $B^{ai}=\delta^{ai}$ is
invariant
under rotations and translations of the configuration space
$\IR^3$ (the spatial
rotations must be combined with a suitably chosen SO(3) gauge
transformation,
constant in this case, as is well known).  Since eq.~(1) is also
covariant,
each such isometry which does not leave $A_i^a$ invariant produces
another
Wu-Yang related potential.  In this way one can extend the potentials
displayed
in (3) to a 4-parameter family, in which the wave has an arbitrary
phase,
$z\to z-z_0$ and direction $(0,0,1)\to \hat{k}$.

\item{c.} Brown and Weisberger [11] are often credited with the proof
that a
uniform non-abelian field strength such as our $B^{ai}$ also requires
a
uniform vector potential, so there is an apparent conflict with the
decidedly
non-uniform potentials displayed in (3).  However Brown and
Weisberger define a uniform field by the strong requirement that
${\underline{\rm all}}$ gauge invariant
quantities, e.g. $B^{ai}D_jB^{ak}$, are constant, so there is no
conflict with their work.

\item{d.} Equation (1) is covariant under diffeomorphisms $y^\alpha
(x^i)$ of
$\IR^3\to \IR^3$ under which $A$ and $B$ transform as follows:
$$
\eqalign{A'^a_\alpha(y) =& {\partial x^i\over\partial
y^\alpha}A_i^a(x) \cr
B'^{a\alpha}(y) =& |{\partial x\over\partial y}|{\partial y^\alpha
\over
\partial x^i}B^{ai}(x) \cr}
\eqno(4)$$
Thus examples of the Wu-Yang ambiguity automatically extend to entire
orbits
of the diffeomorphism group.  The magnetic energy
$$
E={1\over 2}\int d^3x\,\delta_{ij} B^{ai}(x) B^{aj}(x)
\eqno(5)$$
involves the fixed Cartesian metric $\delta_{ij}$, and is not
invariant under
diffeomorphisms.  Below we will show that one can find a
diffeomorphism under
which the field $B^{ai}=\delta^{ai}$ which has infinite
energy, transforms to a configuration $B'^{a\alpha}(y)$ which falls
sufficiently fast as $y^\alpha\to\infty$ that it has finite energy.

2.\quad Let us now review the spatial geometry of [16] which is the
main tool
used in this work.  Let $B^a{}_i(x)$ denote the matrix inverse of the
magnetic
field of SU(2) gauge theory, and $\det B=\det B^{ai}$, which is gauge
invariant.  Then $G_{ij}(x)=B^a{}_i(x)B^a{}_j(x)/\det B$ is a gauge
invariant
symmetric tensor (under diffeomorphisms)\footnote{*}{One sees that
$G_{ij}$ is
a positive definite (negative definite) tensor, when $\det B>0$
$(\det B<0)$  We discuss  the positive case here because only that is
used below.  See [16] for
a complete treatment.}.  The following geometry, which
obviously uses the fact that the gauge group SU(2) is also the
tangent space
group of a 3-manifold, emerged from the physical aim of studying the
action
of the electric field on gauge invariant state functionals $\psi[G]$.

The quantity $b_i^a=|\det B|^{1\over 2}B_i^a$ is essentially a frame
for
$G_{ij}$.  One may apply a Yang-Mills covariant derivative and define
a
quantity $\Gamma_{ij}^k$ as follows:
$$
D_i b_j^a=\Gamma_{ij}^k b_k^a .
\eqno(6)$$
It then follows immediately that $\Gamma_{ij}^k$ transforms as a
connection
under diffeomorphisms and that it is well defined
in terms of $A_i^a$ and $B^{ai}$ for
all non-singular magnetic fields, $\det B\neq0$.  If we multiply (6)
by
$b_k^a$ and symmetrize in $j$, $k$, we easily derive
$$
\partial_i G_{jk}-\Gamma_{ij}^l G_{lk}-\Gamma_{ik}^l G_{jl}=0\ .
\eqno(7)$$
This means that $\Gamma$ is a metric compatible connection for $G$,
and can be
written as
$$
\Gamma_{ij}^k=\dot\Gamma_{ij}^k(G)-{K_{ij}}^k
\eqno(8)$$
where $\dot\Gamma$ is the Christoffel connection and $K$ is the
contortion
tensor, which is antisymmetric in the last pair of indices,
$K_{ijk}=-K_{ikj}$.

Next one studies the anti-symmetrized Yang-Mills derivative of (6)
$$
[D_i,D_j]b_k^a=b_l^a\partial_i\Gamma_{jk}^l+\Gamma_{jk}^l\partial_i
b_l^a -
(i\leftrightarrow j)
\eqno(9)$$
One uses (6) again to find
that the curvature tensor of $\Gamma$ appears.  On the left one uses
the gauge
field Ricci identity and a standard formula for the inverse of a
$3\times3$
matrix.  The result is
$$
R_{kij}^l=\delta_j^l G_{ik}-\delta_i^l G_{jk}\ ,
\eqno(10)$$
the statement that the spatial geometry associated with the gauge
fields $A$,
$B$ is maximally symmetric.  However, in 3 dimensions, with or
without
torsion, the Rieman tensor $R_{kij}^l$ is fully determined by its
Ricci
contraction $R_{kj}=\delta_l^i R_{kij}^l$, so that no information is
lost by
restricting consideration to the contracted form of (8), namely
$$
R_{ij}(\Gamma)=-2G_{ij}
\eqno(11)$$
which defines an Einstein geometry with torsion.  One may show [16]
using the
second Bianchi identity of curvatures with torsion, that an
integrability
condition for (11) is that the contortion tensor is traceless,
${K_{kj}}^k=0$,
and that this is also a direct requirement of the gauge field Bianchi
identity, $D_i B^{ai}=0$, applied to the definition (6) of $\Gamma$.

The discussion above defines the forward map from Yang-Mills fields
$A$ and
$B$, always related by (1), to geometric variables $G$ and $\Gamma$
defined
by explicit local formulas above. The gauge field Ricci and Bianchi
identities
then imply that $G$ and $\Gamma$ are related by the Einstein
condition (11)
with traceless contortion.  The fundamental reason for the Einstein
geometry
is that the magnetic field is simultaneously the curvature (1) of the
gauge
connection $A$ and also essentially the frame of the spatial
geometry.

One may also ask about the inverse map from tensor $G_{ij}(x)$ and
connection
$\Gamma_{ij}^k(x)$ on $\IR^3$ to gauge fields.  Suppose that a frame
$b^a_i$, with $\det b>0$, is constructed for $G_{ij}$ by any standard
method, then (6) can be written out as
$$
\partial_i b_j^a-\Gamma_{ij}^k b_k^a+\epsilon^{abc}A_i^b b_j^c=0\ .
\eqno(12)$$
This is just the ``dreibein postulate'' with $A$ essentially the spin
connection, and one can solve for $A$, obtaining [16]
$$
A_i^a=-{1\over 2}\epsilon^{abc}b^{bj}\left(\partial_i
b_j^c-\Gamma_{ij}^k
b_k^c\right)
\eqno(13)$$
while the magnetic field is defined from the inverse frame by
$$
B^{ai}(x)=|\det G_{jk}|^{1\over 2} b^{ai} .
\eqno(14)$$
Thus given a frame one obtains the magnetic field from (14), while
both $b$
and $\Gamma$ are required to define the potential via (13).  Since
the frame is
unique up to a local $SO(3)$ rotation, these maps define $A$ and $B$
uniquely up to an $SU(2)$ gauge transformation.  Furthermore,
$A$ and $B$ defined in this way satisfy the gauge
theory relation (1) if $\Gamma$ and $G$ satisfy (11).

Thus the gauge theory Wu-Yang ambiguity will appear whenever the
Einstein
equation (11) viewed as a partial differential equation for $K$,
given $G$, has
multiple solutions.  To investigate this it is useful to use the
representation [16]
$$
K^i{}_{jk}=\epsilon_{jkn}S^{ni}{1\over |\det G|^{1/2}}
\eqno(15)$$
which automatically satisfies the antisymmetry
and tracelessness requirements if $S^{ni}$ is a
symmetric tensor.  When (11) is expanded out using (8)
and (15), one finds that the Einstein equation is equivalent to
$$
{\epsilon^{jkl}\over |\det G|^{1/2}}\dot\nabla_k
S_{li}-\left(S_k^j
S_i^k-S_k^k S_i^j\right)=\dot R_i^j+2\delta_i^j .
\eqno(16)$$
In (16) $\dot\nabla_k$ indicates a spatial covariant derivative with
Christoffel connection $\dot\Gamma$ and $\dot R_{ij}$ is the
conventional
symmetric Ricci tensor.  The $\epsilon\dot\nabla S$
term is non-symmetric, so that (16) comprises 9 equations for the 6
components of $S_{ij}$.  However it was shown explicitly in [16] that
there is a Bianchi identity which imposes 3 constraints on the 9
equations, so there is
no reason to think that (16) is an overdetermined system. From (8),
(13) and (15), $A$ can be expressed in terms of $S$ as
$$A_i^a=-{1\over 2}\epsilon^{abc}b^{bj}\dot\nabla_i b^c_j -
b^{ak}S_{ki} .
\eqno(17)$$

Our approach to the Wu-Yang ambiguity is to take an input metric
$G_{ij}(x)$ and study the solutions of (16) for the torsion
$S_{ij}(x)$.
It is not clear why this should be a simpler method than to study
directly
whether (1) has multiple solutions for $A$, given $B$.  Perhaps it is
because
an equation for the 6 components of $S_{ij}$ is simpler to handle
than an
equation for the 9 components of $A_i^a$, but it may just be an
historical
accident that has led us to approach the Wu-Yang ambiguity via the
spatial
geometry of [16].

Before beginning to study applications of (16), it is perhaps useful
to note
that (6) indicates that $\Gamma$ is completely determined by first
covariant
derivatives $D_i B^{aj}$ of the magnetic field.  It then follows from
properties of the inverse map discussed above that there is no
Wu-Yang
ambiguity for the potential $A_i^a$, if we require that both $B$ and
$DB$  are
preserved\footnote{**}{Actually it is sufficient to require that
$D_{[i}b^a_{j]}$ is preserved, because this determines the torsion
tensor from
(6).}. In 4 dimensional $SU(2)$ gauge theory [2,3], the field
strength and its first two covariant derivatives determine the
potential
locally uniquely.

The Wu-Yang ambiguity indicates that the potential $A_i^a(x)$
contains gauge
invariant information beyond that in the magnetic field $B^{ai}(x)$.
Therefore the change of field variable $A_i^a\to B^{ai}$ which was
the basis
of the version of gauge invariant Hamiltonian dynamics presented in
[16] is
invalid.  The discrete 2:1 ambiguity envisaged there could be
handled, but it
is probably impossible to deal with a continuous ambiguity without
serious
revision of the proposal. In the 2+1 Lagrangian theory the equation
of motion
$D^\mu F_{\mu\nu}=0$ provides additional information, presumably
enough to fix the Wu-Yang ambiguity, which then should not be a
difficulty for [17].

3.\quad It is the tensor $G_{ij}(x)=\delta_{ij}$
that corresponds to the magnetic field
$B^{ai}=\delta^{ai}$, and it can be seen without
difficulties that the torsion solutions of (14) are
related in this simple case to the potentials of (3) by
$A^a_i(z)=-S_{ai}(z)$.
 Note that at $\beta=1$, the solution (3) reduces to
$S_{ij}=\delta_{ij}$, and
was already noted in [16].  We found the family of solutions (3) by
first
linearizing about $S_{ij}=\delta_{ij}$ and using Fourier analysis to
find
linearized modes of wave number $k^2=1$.  This led us to investigate
the
single variable ansatz $S_{ij}(z)$ which reduces (16) to a non linear
system
of ordinary differential equations.  Some fiddling then led to (3),
which is
unique within this ansatz (except for translation $z\to z-z_0$).
One can show that the only spherically
symmetric solutions of (16) for input $G_{ij}=\delta_{ij}$ are the
solutions
$S_{ij}=\pm\delta_{ij}$ recognized in [16]. There is a heuristic
argument that
the potentials displayed in (3) together with those obtained from
them by
translation and rotation are the only potentials for the field
$B^{ai}=\delta^{ai}$ which continuously limit to potentials
$\bar A^a_i=\delta^a_i$ with $\beta=1$ in (3). The reason is that one
can
show using the Fourier transform that the set of linear perturbations
about
$\bar A^a_i$ obtained in the $\beta\to 1$ limit of our Wu-Yang family
are complete.

We now discuss the use of diffeomorphisms in relation to the
energy of field configurations exhibiting the Wu-Yang
ambiguity. We proceed in several steps.

\item{a.} Given an initial configuration $B^{ai}(x)$ in
Cartesian coordinates $x^m$ on $\IR^3$, we transform to
spherical coordinates $y^i=(r,\theta,\varphi)$. Both the
spacetime metric $\delta_{ij}$ and the magnetic field $B^{ai}(x)$
transform to
$$ \eqalign{\delta_{ij} \to & g_{ij}(y)={\partial x^m\over \partial
y^i}
{\partial x^n\over \partial y^j} \delta_{mn},\cr
B^{ai} \to & B'^{ai}(y)=|{\partial x\over \partial y}|
{\partial y^i\over \partial x^m} B^{am}(x),\cr} \eqno(18) $$
and the energy (5) can be rewritten as

$$ E={1\over 2} \int d^3y {1\over \sqrt{g(y)}}g_{ij}(y)B'^{ai}B'^{aj}
={1\over 2} \int d^3y {1\over \sqrt{g(y)}}g_{ij}(y)G'^{ij}(y)det G'.
  \eqno(19) $$
For the specific case of spherical coordinates and
field $B^{ai}=\delta^{ai}$ this is equal to the divergent expression
$$ E={3\over 2} \int dr d\theta d\varphi r^2 \sin\theta , \eqno(20)
$$
which could have been obtained by naive change of variables in (5).

\item{b.} We now make the diffeomorphism $y^i\to z^i$ of
the vector potential and the magnetic field by changing the radial
coordinate
$$ \eqalign{z^1&=R(r),\cr z^2&=\theta=y^2,\cr z^3&=\varphi=y^3,\cr
\tilde{B}^{ai}&=|{\partial y\over \partial z}|{\partial z^i\over
\partial y^m}
B'^{am}(y),\cr
\tilde{G}^{ij}&={\partial z^i\over \partial y^m}
{\partial z^j\over \partial y^n} G'^{mn}(y).\cr} \eqno(21) $$
In the specific case $B^{ai}=\delta^{ai}$, $G'^{mn}$
is the standard diagonal inverse metric for $(r,\theta,\varphi)$,
and $\tilde{G}^{ij}$ is also diagonal and given by
$$ \eqalign {\tilde{G}^{RR}&=R'^2,\cr \tilde{G}^{\theta
\theta}&={1\over r^2},\cr \tilde{G}^{\varphi\varphi}&=
{1\over r^2\sin^2\theta},\cr} \eqno(22) $$

\item{c.} The coordinates $z^\alpha =(R,\theta,
\varphi)$ are new spherical coordinates on $\IR^3$,
so the energy of the $\tilde{B}^{ai}(z)$ field configuration is
$$  \tilde{E}={1\over 2} \int d^3y {1\over \sqrt{g(z)}}g_{ij}(z)
\tilde{G}^{ij}(z)det \tilde{G}(z),
  \eqno(23) $$
with $g_{ij}$ the standard metric for $(R,\theta,\varphi)$. For the
example
$B^{ai}=\delta^{ai}$, one can use (18-23) and obtain
$$ \tilde{E}={1\over 2} \int dR d\theta d\varphi {r^4\sin\theta\over
R^2}
\left(1+{2\over R'^2}\right).  \eqno(24) $$
The constraints that $\tilde{B}^{ai}(z)$ is a finite
energy field configuration on $\IR^3$ are that $R(r)$ map
$(0,\infty) \to (0,\infty)$ monotonically such that $r^4/R^2$
falls faster than $1/R$ and no other singularities appear. It
is not difficult to show that $R=r\sqrt{1+r^8}$ satisfies these
requirements.
It is certainly not unique.

4. The next set of Wu-Yang examples we discuss emerge from the
3-dimensional hyperbolic metrics
$$ ds^2={1\over c^2z^2}\left( dx^2 + dy^2 + dz^2\right) \eqno(25) $$
for which $R_{ij}=-2c^2 G_{ij}$. One may anticipate that the case
$c^2=1$ is
especially simple because the right side of (16) vanishes. It turns
out that
one can also make the $\epsilon \dot\nabla S$ and $SS$ terms vanish
separately.
There is no integrability constraint when $\dot\nabla_j$ is applied
to the former condition, while the second condition implies that
$S_{ij}$ is a rank 1
dyadic matrix. With this structure in view one can easily find that
within
the two variable ansatz $S_{ij}(z,x)$, there is the family of
solutions
$$ S_{ij}(z,x)=\delta_{i1} \delta_{j1} {1\over z}h(x) \eqno(26) $$
which involves an arbitrary function of the variable $x$. The
solution can
be rotated by an angle $\theta$ in the $x,y$ plane to obtain
$$ \eqalign{S_{ij}&={1\over z} h(x\cos\theta+y\sin\theta) V_i V_j
,\cr
V_i&=(\cos\theta,\sin\theta,0).\cr} \eqno(27) $$
We have not studied the application of the full $SO(2,1) \times
SO(2,1)$
isometry group of the metric (25), but more solutions seem likely. In
this
frame the magnetic field is given simply by $B^{ai}=\delta^{ai}/z^2$
while the   gauge potential corresponding to (26) is obtained from
equation (17):
$$ A_1^1=-h(x), \quad A_2^1=-A_1^2=1/z, \eqno(28)$$
with the rest of the components vanishing. In this frame, however,
the magnetic
field $B^{ai}$ is singular on the plane $z=0$, so we transform our
configuration to a frame in which both
the magnetic field and gauge potential are manifestly regular over
all of
$\IR^3$.
Transforming to the coordinates
$$ \eqalign{u^1&={4x\over x^2+y^2+(z+1)^2},\cr
u^2&={4y\over x^2+y^2+(z+1)^2},\cr
u^3&={2(x^2+y^2+z^2-1)\over x^2+y^2+(z+1)^2},\cr} \eqno(29) $$
the metric can be re-written in the form
$$ ds^2={1\over (1-r^2/4)^2}\left((du^1)^2+
(du^2)^2+(du^3)^2\right), \eqno(30) $$
where $r^2=(u^1)^2+(u^2)^2+(u^3)^2 < 4$, which is another standard
metric on a $3$-hyperboloid and can be analytically continued to a
metric
on a $3$-sphere. The inverse transformation is then given by
$$ \eqalign{x&={u^1\over (u^1/2)^2+(u^2/2)^2+(u^3/2-1)^2},\cr
 y&={u^2\over (u^1/2)^2+(u^2/2)^2+(u^3/2-1)^2},\cr
z&={1-((u^1/2)^2+(u^2/2)^2+(u^3/2)^2)\over
(u^1/2)^2+(u^2/2)^2+(u^3/2-1)^2}.\cr} \eqno(31) $$
We make a further transformation to the coordinates
$$ v^i=u^i/(1-r^2/4) \eqno(32)$$
 so that the entire $\IR^3$ is covered. The inverse transformation is
given by
$$ u^i={2(\sqrt{1+v^2}-1)\over v^2} v^i \eqno(33)$$
so that the metric can be written in the form
$$ ds^2={dv^2\over 1+v^2} + dv^2d\Omega^2=dv^idv^i-
({1\over 1+v^2})v^idv^iv^jdv^j. \eqno(34)$$
{}From the last form, one can define a frame and use (14) to obtain
the
following regular configuration for the $B-$field:
$$ B^{ai}={\delta^{ai}\over \sqrt{1+v^2}} + \left(1-{1\over
\sqrt{1+v^2}}\right) {v^av^i\over v^2} , \eqno(35) $$
in which $v^i$ are regarded as Cartesian coordinates.
A dyadic solution for $S$ in the $(xyz)$ frame then transforms to a
dyadic solution in the $(v^1v^2v^3)$ frame according to
$$ S'_{ij}(v^k)=S_{mn}(x^p){\partial x^m
\over \partial v^i}{\partial x^n\over \partial v^j}.\eqno(36) $$
In particular, the solution $S_{11}(x,z)=h(x)/z$ transforms into
$$ S'_{ij}(v^k)={1\over z(v^k)} h(x(v^k)) V_{(i)} V_{(j)}, \eqno(37)
$$
where $V_{(i)}=\partial x/\partial v^i$ and where $x(v^k)$ and
$z(v^k)$ are easily given from (31) and (33). Using equation (17) we
compute the
corresponding set of potentials $A^a_i$ in the $v^k$-frame all of
which yield the magnetic field of (35):
$$ A^a_i=-\epsilon^{aik}\left({\sqrt{1+v^2}-1\over v^2}\right)v^k -
S'_{ai}-\left({\sqrt{1+v^2}-1\over v^2}\right)v^a v^k S'_{ki}
.\eqno(38) $$

When $c^2\neq 1$ the full nonlinear equations are very difficult to
handle, so
we restrict ourselves to a perturbative expansion about the symmetric
solution
$\bar S_{ij}=\sqrt{1-c^2}G_{ij}$ of (16) by setting $S_{ij}=\bar
S_{ij} +
\hat\Sigma_{ij}$. The perturbation $\hat\Sigma_{ij}$ satisfies the
linear equation
$$ (cz) \epsilon^{jkl}\dot\nabla_k \hat\Sigma_{li} +
\sqrt{1-c^2}(\hat\Sigma_{ji} +
\hat\Sigma_{kk}\delta_{ji})=0 .\eqno(39) $$
(the placement of the $j$ index reflects the removal of the conformal
factor.)
The 9 equations for the 6 components of $\hat\Sigma$ cannot all be
independent,
and the fact that we find a consistent solution below is a practical
test
of the exact Bianchi identity [16] satisfied by (16). Note that the
$ij$
contraction of (39) immediately tells us that the trace
$\hat\Sigma_{kk}=0$.

Because of the $x$-translation symmetry of the metric (25) we look
for a solution
of the form $\hat\Sigma_{ij}(z,x)=\Sigma_{ij}(z,k) e^{ikx}$. The 9
equations of
(39) can be manipulated to obtain a second order differential
equation for the
component $\Sigma_{23}$
$$ \left(z^2 {d^2\over dz^2} + z {d\over dz} - k^2z^2 + {1-c^2\over
c^2}\right)
\Sigma_{23}=0 ,\eqno(40) $$
while the other components are related to $\Sigma_{23}$ by
$$ \eqalign{ \Sigma_{33}&={-ikc\over \sqrt{1-c^2}}z \Sigma_{23} ,\cr
\Sigma_{13}&={c\over \sqrt{1-c^2}}z {d\over dz}\Sigma_{23} ,\cr
\Sigma_{12}&={i\over k}({d\over dz} - {1\over z})\Sigma_{23} ,\cr
\Sigma_{11}&={-ic\over k\sqrt{1-c^2}}z {d^2\over dz^2}\Sigma_{23}
,\cr
\Sigma_{22}&=-(\Sigma_{11} + \Sigma_{33}) .\cr } \eqno(41) $$
Note that (40) is the differential equation for Bessel functions of
imaginary
argument $ikz$ and index $p=\sqrt{(c^2-1)/c^2}$ which is also
imaginary when
$c^2<1$ and the symmetric torsion $\bar S_{ij}$ is real.

Note that the wave number $k$ of the linear perturbation is not
restricted
in contradistinction to the flat metric where, as can be seen from
the small
amplitude limit $\beta\to 1$ in (3), the wave number $k=1$ is
required. This
means that the general real superposition
$$ \int dk \varphi(k) \Sigma_{ij}(kz)e^{ikx} + {\rm c.c.} \eqno(42)
$$
is also a solution, so we have the freedom of an arbitrary function
at
the linear level. We expect that (42) can be used as the ``input'' to
the
system of differential equations determining second and higher order
perturbative solutions of (16), and that the functional freedom of
$\varphi(k)$
remains. Thus the qualitative picture of the torsion solutions for
the
hyperbolic metrics for all values of $c$ is that they contain an
arbitrary
function of a single variable and the additional parametric freedom
obtained
from isometries. The case $c=1$ is special only beause exact
solutions can be
easily obtained.

5. The final set of Wu-Yang ambiguities we discuss arise from the
$2+1$
product metrics of the form
$$ ds^2=F(x,y)(dx^2+dy^2) + dz^2, \eqno(43) $$
and for which $\dot R_{ij}=-\delta_{ij}\nabla^2 \ln F$ and
$\dot R_{i3}=\dot R_{33}=0$
for $i,j=1,2$. Then provided $F$ satisfies the differential equation
$$ \nabla^2 \ln F = c^2F + {m^2\over F}, \eqno(44) $$
where $c^2<1$ and $m>0$, a solution for $S_{ij}$ is given by
$$
S_{ij}=\pmatrix{\beta F\pm m \cos (z/\beta) &\pm m sin(z/\beta) & 0
\cr
               \pm m sin(z/\beta) &\beta F \mp m cos(z/\beta),  & 0
\cr
	       0 & 0 & 1/\beta \cr}
\eqno(45)$$
where $\beta=\sqrt{1-c^2}$. For $m\neq 0$, this
solution represents a continuous 1-parameter family of
Wu-Yang ambiguities obtained by translations
in $z$.

For $m=0$ the metric (43) reduces to the constant curvature case,
the wave disappears, and the torsion becomes
$$ \tilde S_{11}= \tilde S_{22}=\beta F,\quad\tilde S_{33}=1/\beta
.\eqno(46)$$
The manifestly symmetric form of the product metric is
$$ ds^2={1\over c^2y^2}\left( dx^2 + dy^2 \right)+ dz^2 ,\eqno(47) $$
and we now study the perturbative expansion about the symmetric
$\tilde S_{ij}$ of (46) by setting $S_{ij}=\tilde S_{ij} +
\tilde\Sigma_{ij}$. Because of the $z$-translation symmetry of the
metric (47) we look for a solution of the form
$\tilde\Sigma_{ij}(y,z)=\Sigma'_{ij}(y,k) e^{ikz}$. The 9 equations
in the linearized approximation can be manipulated to obtain a second
order differential equation for the
component $\Sigma'_{23}$
$$ \left( {d^2\over dy^2} + {2\over y} {d\over dy} +
{\alpha(a,k)\over y^2} \right) \Sigma'_{23}=0 ,\eqno(48) $$
where $\alpha(\beta,k)=2\beta^2(\beta^2-k^2)/(1-\beta^4)$ while the
other components are simply related to $\Sigma'_{23}$. The solution
of (48) is given simply by the power behaviour $\Sigma'_{23}=A y^t$,
where $A$ is an arbitrary
constant and $t=-1/2 \pm \sqrt{1/4 - \alpha(\beta,k)}$. For
$\tilde\Sigma_{ij}(y,x)=\Sigma''_{ij}(y,k) e^{ikx}$, $\Sigma''_{ij}$
is simply expressed in terms of Bessel functions of imaginary
argument  $iky$ and index
$m=\sqrt{1/4 - 2\beta^2/(1-\beta^4)}$, which is either real or
imaginary depending on the value of $\beta$.

Again the wave number $k$ of the linear perturbation is not
restricted
and we expect that this functional freedom persists in higher order
perturbative solutions.

6. What we have exhibited in this paper are several examples of a
continuous
Wu-Yang ambiguity for $SU(2)$  gauge fields in 3 dimensions and a new
technique,
namely the Einstein space condition (14) for obtaining such field
configurations. It is intriguing to ask about the systematics of the
ambiguity;
namely what properties of the $B$-field determine the degree of
ambiguity in the
associated potentials $A$. Our examples provide at least a limited
view of this
systematics. Certainly an ambiguity is generated whenever there is a
symmetry
transformation of $B$ which acts nontrivially on $A$, but this is not
enough to
explain the free parameter $\beta$ in (3), nor the arbitrary
functions such as
$F(x)$ in the example (26-38) or in the various linear solutions we
have
presented. Gauge field topology does not seem to be the issue here
for two
reasons. First of all the ambiguity can be exhibited in any compact
subset
of the configuration space $\IR^3$. Second, if we are given in some
gauge a
Wu-Yang family with suitable behavior at spatial infinity one can
apply a
gauge transformation to change the topological class at will. Of
course
one does expect that, except for singularities of the map (1), the
degree
of ambiguity in $A$ will not change as the parameters of $B$ are
smoothly varied.
Our examples appear to be consistent with this requirement, although
the
discrete ambiguity found when $\dot R_{ij}=0$ must be understood as a
singular limit
of the case of non-zero curvature. It is interesting that the
Riemannian
curvature $\dot R_{ij}$ of the metric $G_{ij}$ obtained from $B$
plays a role both in the
ease of obtaining solutions for $A$ and in the qualitative nature of
the
ambiguity.

\centerline{{\bf Acknowledgements}}
We wish to thank L. Alvarez-Gaum\'e, M. Bauer, S. Deser, C. Imbimbo, J.
Liu
and E. Verlinde for helpful discussions.
\par
\vfill\eject

\centerline{{\bf References}}
\bigskip
\baselineskip 20pt plus 1pt minus 1pt

\item{[1]} T.T. Wu and C.N. Yang, {\it Phys. Rev.} {\bf D12} (1975)
3845.

\item{[2]} C.H. Gu and C.N. Yang, {\it Sci. Sin.} {\bf 18} (1975)
484;
{\it Sci. Sin.} {\bf 20} (1977) 47.

\item{[3]} C.L. Shen, {\it Fudan Journal} (Natural Science) {\bf 2}
(1976) 61.

\item{[4]} S. Deser and F. Wilczek, {\it Phys. Lett.} {\bf 65B}
(1976) 391.

\item{[5]} R. Roskies, {\it Phys. Rev.} {\bf D15} (1977) 1731.

\item{[6]} M. Calvo, {\it Phys. Rev.} {\bf D15} (1977) 1733.

\item{[7]} S. Coleman, {\it Phys. Lett.} {\bf 70B} (1977) 59.

\item{[8]} M.B. Halpern, {\it Phys. Rev.} {\bf D16} (1977) 1798;
{\it Nucl. Phys.} {\bf B139} (1978) 477.

\item{[9]} S. Solomon, {\it Nucl. Phys.} {\bf B147} (1979) 174.

\item{[10]} C.G. Bollini, J.J. Giambiagi and J. Tiomno, {\it Phys.
Lett.}
{\bf 83B} (1979) 185.

\item{[11]} L.S. Brown and W.I. Weisberger, {\it Nucl. Phys.} {\bf
B157}
(1979) 285.

\item{[12]} S. Deser and W. Drechsler, {\it Phys. Lett.} {\bf 86B}
(1979) 189.

\item{[13]} M. Mostow, {\it Commun. Math. Phys.} {\bf 78} (1980) 137.

\item{[14]} F.A. Doria, {\it Commun. Math. Phys.} {\bf 79} (1981)
435;
{\it J. Math. Phys.} {\bf 22} (1981) 2943.

\item{[15]} A.M. Badalyan, {\it Sov. J. Nucl. Phys.} {\bf 38} (1983)
464.

\item{[16]} D.Z. Freedman, P.E. Haagensen, K. Johnson and J.I.
Latorre,
CERN-TH.7010/93, CTP \#2238, hep-th 9309045 (September 1993).

\item{[17]} F.A. Lunev, {\it Phys. Lett.} {\bf B295} (1992) 99.

\end